\documentclass[preprint,3p,times,onecolumn]{elsarticle}
\usepackage{epsfig}
\usepackage{subfig}
\usepackage{amssymb}
\journal{Physics Letters A}
\begin{document}
\begin{frontmatter}

\title{Exact behavior of the energy density inside a one-dimensional oscillating cavity with a thermal state}

\author[ufpa]{Danilo T. Alves}
\ead{danilo@ufpa.br}

\author[cbpf,ufpa3]{Edney R. Granhen}
\ead{edney@ufpa.br}

\author[ufpa]{Hector O. Silva \corref{hosilva} \fnref{fn1}}
\ead{hosilva@ufpa.br}

\author[ufpa3,ufpa2]{Mateus G. Lima}
\ead{mateusl@ufpa.br}

\cortext[hosilva]{Corresponding author}
\fntext[fn1]{Tel/Fax: +55 91 3201-7430 (Brazil)}

\address[ufpa]{Faculdade de F\'\i sica, Universidade Federal do Par\'a, 66075-110, Bel\'em, PA,  Brazil}
\address[cbpf]{Centro Brasileiro de Pesquisas F\'\i sicas, Rua Dr. Xavier Sigaud, 150, 22290-180, Rio de Janeiro, RJ, Brazil}
\address[ufpa3]{Faculdade de Ci\^encias Exatas e Naturais, Universidade Federal do Par\'a, 68505-080, Marab\'a, PA,  Brazil}
\address[ufpa2]{Faculdade de Engenharia El\'etrica, Universidade Federal do Par\'a, 66075-110, Bel\'em, PA,  Brazil}

\begin{abstract}
We investigate the exact behavior of the energy density of a real massless scalar field inside a cavity with a single moving mirror executing a resonant oscillatory law of motion, satisfying Dirichlet boundary conditions at finite temperature. Our results are compared with those found in literature through analytical approximative methods.
\end{abstract}

\end{frontmatter}

\section{Introduction}
\label{Introduction}
Since the pioneering work of Moore \cite{Moore-1970} and the subsequent works published in the 70s 
(for instance Refs. \cite{Fulling-Davies-PRS-1976-I, Fulling-Davies-PRS-1977-II, trabalhos-pioneiros}), the problem of particle creation from vacuum due the interaction of a quantum field with moving mirrors (dynamical Casimir effect) has been subject of intense 
theoretical study (for recent reviews see Refs. \cite{Reviews} and references therein). 
On the experimental side, the dynamical Casimir effect has not yet been observed, despite some experimental schemes have been proposed
based on simulation of a moving mirror by changing the reflectivity of a semiconductor using laser beams \cite{Experimento-MIR} or, more recently, using a coplanar waveguide terminated by a superconducting quantum interference device \cite{Johansson-PRL-2009}.

In Ref. \cite{Moore-1970}, Moore considered a real massless scalar field in a two-dimensional spacetime inside a cavity with one moving boundary imposing Dirichlet boundary conditions on the field. He obtained the field solution in terms of 
the so-called Moore's equation. Exact analytical solutions for particular laws of motion of the boundary \cite{exact-analytical-solutions},
and also approximative analytical solutions \cite{Dodonov-JMP-1993,Dalvit-PRA-1998}
for the Moore's equation were obtained but, so far, there is no general technique to find analytical solutions for it. 
On the other hand, Cole and Schieve developed a geometrical approach to solve numerically
and exactly Moore's equation \cite{Cole-Schieve-1995}, obtaining
the exact behavior of the energy density in a non-stationary cavity 
considering vacuum as the initial field state \cite{Cole-Schieve-1995,Cole-Schieve-2001}. 
The dynamical Casimir effect was also studied with different approaches from that adopted by Moore,
via perturbative methods for a single mirror \cite{single-mirror-I} and for
an oscillating cavity \cite{cavity}.

The quantum problem of moving mirrors with initial field states different from 
vacuum was also analyzed \cite{Fulling-Davies-PRS-1977-II,temperatura-uma-fronteira, temperatura-cavidade, 
estados-coerentes, Dodonov-Andreata-JPA-2000}.
Specifically, a thermal bath 
was investigated for the case of a single mirror 
\cite{temperatura-uma-fronteira},
and also for an oscillating cavity \cite{temperatura-cavidade,
Dodonov-Andreata-JPA-2000}.
In a recent paper \cite{alves-granhen-silva-lima-2010}, the present authors 
investigated the behavior of the energy density 
inside a cavity with a moving mirror,
for an arbitrary initial field state,
obtaining formulas which enabled us to get
exact numerical results for the quantum radiation force and for 
the energy density in a non-static cavity for an arbitrary initial field state and law of motion for the moving boundary.
However, in Ref. \cite{alves-granhen-silva-lima-2010} 
we applied our formulas only to the case of an expanding cavity, with a non-oscillating movement
and relativistic velocities.
In the present letter, we consider the exact approach 
developed in Ref. \cite{alves-granhen-silva-lima-2010} to 
study the behavior of the energy density for a non-static cavity with a thermal bath and a
resonant oscillatory law of motion.
We compare our results with those found
by Andreata and Dodonov in Ref. \cite{Dodonov-Andreata-JPA-2000}
where this problem was investigated through analytical approximative methods for
small oscillations. We show the limitations in the results
obtained via this approach as the amplitude of oscillation grows to
outside the small oscillations regime. We also investigate the energy density outside this regime for
the first time. Additionally, we show that for small amplitudes of oscillation our results are in excellent
agreement with \cite{Dodonov-Andreata-JPA-2000}.  

This Letter is organized as follows: in Sec. \ref{exact-formulas} we obtain the field solution and the exact formulas for the energy density taking the initial field state as a thermal bath and considering that the mirrors impose Dirichlet boundary condition on the field. 
In Sec. \ref{behavior} we study the behavior of the energy density for the case of an oscillatory law of motion for the moving boundary. Finally, in Sec. \ref{conclusions} we summarize our results.
\section{Exact formulas for the energy density}
\label{exact-formulas}
Let us start considering the real massless scalar field satisfying the Klein-Gordon equation
(we assume throughout this paper $\hbar=c=k_B=1$):
$
\left(\partial _{t}^{2}-\partial _{x}^{2}\right) \psi \left(
t,x\right) =0,
$
and obeying Dirichlet boundary conditions imposed at the static boundary located at $x=0$, and also at the moving boundary's position at $x=L(t)$, that is
$\psi\left( t,0\right)=\psi\left( t,L(t)\right)=0$, where $L(t)$ is an arbitrary prescribed law for the moving boundary with $L(t\leq 0)=L_0$, with $L_0$ being the length of the cavity in the static situation.
%

The field in the cavity can be obtained
by exploiting the conformal invariance of the Klein-Gordon equation \cite{Moore-1970, Fulling-Davies-PRS-1976-I}. 
The field operator, solution of the wave equation, can be written as:
\begin{equation}
\hat{\psi}(t,x)=\sum^{\infty }_{n=1}\left[
\hat{{a}}_{n}\psi_{n}\left( t,x\right) +H.c.\right],
\label{field-solution-1}
\end{equation} 
where the field modes $\psi_{n}(t,x)$ are given by:
\begin{eqnarray}
\psi_{n}(t,x)=\frac{1}{\sqrt{4n\pi}}\left[\;\varphi_{n}(v)
+ \;\varphi_{n}(u)\right],
\label{field-solution-2}
\end{eqnarray} 
with $\varphi_{n}(z)=\exp\left[{-in\pi R(z)}\right]$, $u=t-x$, $v=t+x$. The function $R$ satisfies Moore's equation:
\begin{equation}
R[t+L(t)]-R[t-L(t)]=2.
\label{Moore-equation}
\end{equation}
Considering the Heisenberg picture, we are interested in 
the averages $\langle...\rangle$ taken over a thermal state. For this particular field state we have
$\langle \hat{{a}}^{\dagger}_{n}\hat{a}_{n'}\rangle=\delta_{nn'}\xi_{n}(T)$
and $\langle \hat{a}_{n}\hat{a}_{n^\prime}\rangle=\langle \hat{{a}}^{\dagger}_{n}\hat{a}^{\dagger}_{n'}\rangle=0$, where 
$
\xi_{n}\left(T\right)=\left\{ \exp{\left[n\pi / \left( L_{0} T\right)\right]} - 1\right\}^{-1}
$
and $T$ is the temperature.

Taking the expected value of the energy density operator ${\cal T}=\langle\hat{T}_{00}(t,x)\rangle$, where \cite{Fulling-Davies-PRS-1976-I}
\begin{equation}
\hat{T}_{00}(t,x)=\frac{1}{2}\left[ \left(\frac{\partial \hat{\psi}}{\partial t}  \right)^{2}+\left( \frac{\partial \hat \psi}{\partial x} \right)^{2} \right],
\end{equation}
we can split the renormalized energy density ${\cal T}$ as follows \cite{alves-granhen-silva-lima-2010}:
\begin{equation}
{\cal T}={\cal T}_{\mbox{\footnotesize vac}}+ {\cal T}_{\mbox{\footnotesize non-vac}}, 
\label{T}
\end{equation}
where
\begin{equation}
{\cal T}_{\;\mbox{\footnotesize vac}} = -f(v) -f(u), 
\label{T-vac-ren}
\end{equation}
\begin{equation}
{\cal T}_{\;\mbox{\footnotesize non-vac}}=-g(v) - g(u),
\label{T-non-vac}
\end{equation} 
with
\begin{equation}
g(z)=-\frac{\pi}{2} \sum_{n=1}^{\infty }n \left[R^{\prime }\left(z\right)\right]^{2}\xi_{n}(T),
\label{g1}
\end{equation}
\begin{equation}
f(z)=\frac{1}{24\pi}\left\{\frac{R^{\prime\prime\prime}(z)}{R^{\prime}(z)}-\frac{3}
{2}\left[\frac{R^{\prime\prime}(z)}{R^{\prime}(z)}\right]^{2}+\frac{\pi
^{2}}{2}{R^{\prime}(z)}^{2}\right\}. 
\label{T-vac-f}
\nonumber	
\end{equation}
In Eqs. (\ref{g1}) and (\ref{T-vac-f}) the derivatives, denoted by the primes, are taken with respect to the argument of the function $R$.

For further analysis, it is useful to write \cite{alves-granhen-silva-lima-2010}:
\begin{equation}
{\cal T}=-h(v)-h(u), 
\label{T-em-termos-h}
\end{equation}
where $h(z)=f(z)+g(z)$.

In the static situation $t\leq 0$, where both boundaries are at rest, the function $R$ is given by $R(z)={z}/{L_0} $\cite{Moore-1970}.
The functions
$f$ and $g$, now relabeled, respectively, as
$f^{(s)}$ and $g^{(s)}$, are now given by:
\begin{equation}
f^{(s)}=\frac{\pi}{{48}L_0^2},
\label{f-static}
\end{equation}
\begin{equation}
g^{(s)}=-\frac{\pi}{2L_{0}^{2}} \sum_{n=1}^{\infty }n\;\xi_{n}(T).
\label{g-static}
\nonumber
\end{equation}
Note that in the static situation ${\cal T}_{\;\mbox{\footnotesize vac}}$ is 
the Casimir energy density ${\cal T}_{\mbox{\footnotesize cas}}=-{\pi}/({24L_0^2})$.
In Ref. \cite{alves-granhen-silva-lima-2010} it was shown that the behavior of the energy density in a cavity is determined by the function $h$, which obeys:
\begin{eqnarray}
h\left[t+L\left( t\right) \right] &=&h\left[t-L\left(
t\right) \right]{\cal{A}}(t)+{\cal{B}}(t).
\label{h}
\end{eqnarray}
where
\begin{equation}
{\cal{A}}(t)=\left[ \frac{1-L^{\prime }\left(
t\right) }{1+L^{\prime }\left( t\right) }\right]^{2},
\label{f-A}
\end{equation}
\begin{eqnarray}
{\cal{B}}(t)&=&-\frac{1}{12\pi }\frac{L^{\prime \prime \prime }\left(
t\right) }{\left[ 1+L^{\prime }\left( t\right) \right]^{3}\left[
1-L^{\prime }\left( t\right) \right] }\nonumber\\
&&-\frac{1}{4\pi }\frac{L^{\prime \prime 2}\left( t\right) L^{\prime }\left(
t\right) }{\left[ 1+L^{\prime }\left( t\right) \right]^{4}\left[
1-L^{\prime }\left( t\right) \right]^{2}}.
\label{f-B}
\end{eqnarray}
Eq. (\ref{h}) enables us to obtain recursively the value of $h(z)$, and consequently the energy density ${\cal{T}}$ (\ref{T-em-termos-h}), in terms of its 
static value 
$$h^{(s)}=f^{(s)}+g^{(s)}.$$
Solving recursively the Eq. (\ref{h}), as discussed in Ref. \cite{alves-granhen-silva-lima-2010} in details, we can write $h(z)$ in the following manner:
\begin{eqnarray}
h(z)&=&h^{(s)}{\cal{\widetilde{A}}}(z)+{\cal{\widetilde{B}}}(z),
\label{h-final}
\end{eqnarray}
where:
\begin{equation}
{\cal{\widetilde{A}}}(z)=\prod_{i=1}^{n(z)}{\cal{A}}[t_{i}(z)],
\label{a-tilde}
\end{equation}
\begin{equation}
{\cal{\widetilde{B}}}(z)=\sum_{j=1}^{n(z)}{\cal B}[t_{j}(z)]\prod_{i=1}^{j-1}{\cal A}[t_{i}(z)],
\label{b-tilde}
\end{equation}
\begin{eqnarray}
z&=&t_1+L(t_1),
\label{z1}
\\ 
t_{i+1}+L(t_{i+1})&=&t_i-L(t_i),\;i=1,2,3...,
\label{z}
\end{eqnarray}
being $n$ the number of reflections of the null lines on the moving boundary world line during the recursive process.
We remark that in Eq. (\ref{h-final}), the functions ${\cal\widetilde{A}}$ and ${\cal\widetilde{B}}$ depend only on the 
law of motion of the moving mirror, whereas the dependence on the initial field state is stored in the static value $h^{(s)}$.
We also observe that, for a generic law of motion, 
${\cal{\widetilde{A}}}$ and ${\cal{\widetilde{B}}}$ are different functions,
with the following properties \cite{alves-granhen-silva-lima-2010}: 
\begin{equation}
{\cal{\widetilde{A}}}(z)>0\;\forall\;z,\;
{{\cal{\widetilde{A}}}(z< L_0)}=1,\;
{\cal{\widetilde{B}}}(z< L_0)=0,
\label{A-b-cond}
\nonumber
\end{equation}

The energy densities ${\cal T}_{\mbox{\footnotesize vac}}$ and ${\cal T}_{\mbox{\footnotesize non-vac}}$ (Eqs. (\ref{T-vac-ren}) and (\ref{T-non-vac})) are now respectively rewritten as: 
\begin{equation}
{\cal T}_{\mbox{\footnotesize vac}}=-f^{(s)}\left[{\cal{\widetilde{A}}}(u)+{\cal{\widetilde{A}}}(v)\right]
-{\cal{\widetilde{B}}}(u)-{\cal{\widetilde{B}}}(v),
\label{T-vac-A-B}
\end{equation}
\begin{equation}
{\cal T}_{\mbox{\footnotesize non-vac}}=-g^{(s)}\left[{\cal{\widetilde{A}}}(u)
+{\cal{\widetilde{A}}}(v)\right].
\label{T-non-vac-A}
\end{equation}
From Eqs. (\ref{T}), (\ref{T-vac-A-B}) and (\ref{T-non-vac-A}) the exact formula for the total energy density  $\cal{T}$ is now given by:
\begin{equation}
{\cal T}=-h^{(s)}\left[{\cal{\widetilde{A}}}(u)+{\cal{\widetilde{A}}}(v)\right]
-{\cal{\widetilde{B}}}(u)-{\cal{\widetilde{B}}}(v).
\label{T-total}
\end{equation}

In this two-dimensional model the instantaneous force $\cal F$ acting on the moving boundary (disregarding the contribution
of the field outside the cavity) 
is given by ${\cal F}(t)={\cal T}[t,L(t)]$. We also define 
${\cal F}_{\mbox{\footnotesize vac}}(t)={\cal T}_{\mbox{\footnotesize vac}}[t,L(t)]$ and
${\cal F}_{\mbox{\footnotesize non-vac}}(t)={\cal T}_{\mbox{\footnotesize non-vac}}[t,L(t)]$.

With this results in hand, we are ready to investigate the exact  behavior of the energy density for a thermal state and make comparison with the results found through the analytical approximative approach found in the literature.
\section{Comparing exact and approximate results for the energy density}
\label{behavior}

From Eqs. (\ref{f-A}), (\ref{f-B}), (\ref{a-tilde}) and (\ref{b-tilde})
we see that the functions ${\cal{\widetilde{A}}}$ and ${\cal{\widetilde{B}}}$ 
are different one from another for an arbitrary law of motion.
Therefore, our first conclusion is that the functions ${\cal T}_{\mbox{\footnotesize vac}}$ and ${\cal T}_{\mbox{\footnotesize non-vac}}$ 
(given by Eqs. (\ref{T-vac-A-B}) and (\ref{T-non-vac-A})) consequently have, in general, 
different structures as well. Hereafter we use the word structure in the following sense:
two graphs have the same structure if they have the same number
of maximum and minimum points and these points are at the same positions in both graphs. 
As a direct consequence of our first conclusion, we can say that the thermal force ${\cal F}_{\mbox{\footnotesize non-vac}}$  
and the radiation reaction force ${\cal F}_{\mbox{\footnotesize vac}}$ 
have in general different structures. At a first glance, our conclusion contrasts with that found 
by Andreata and Dodonov \cite{Dodonov-Andreata-JPA-2000}, according to which ${\cal T}_{\mbox{\footnotesize vac}}$ and ${\cal T}_{\mbox{\footnotesize non-vac}}$  exhibit a same structure for initial diagonal states (as the case of the thermal state). Next we will discuss this issue.

Let us consider the particular laws of motion given by
\begin{equation}
L(t)=L_{0}\left[ 1+\epsilon\sin\left( \frac{p\pi t}{L_{0}} \right) \right],
\label{law-of-motion}
\end{equation}
where $L_{0}=1$, $p=1,2,...$, 
and $\epsilon$ is a dimensionless parameter.  
This oscillatory boundary motion was investigated in several papers (see, for instance, Refs. \cite{Dodonov-JMP-1993,Dodonov-Andreata-JPA-2000}), with the calculation of the energy density developed in the context of analytical approximative methods,
considering small amplitudes of oscillation ($\left\vert \epsilon \right\vert \ll 1$). 
Taking as basis the results found in Ref. \cite{Dodonov-Andreata-JPA-2000},
the renormalized energy density $\cal T$, corresponding to the laws of
motion (\ref{law-of-motion}) is given by  ${\cal T}\approx{\cal T}^{(a)}$, with
\begin{equation}
{\cal T}^{(a)}=-(h^{(s)}-p^2 f^{(s)})[s(u)+s(v)]- 2 p^2 f^{(s)},
\label{T-a}
\end{equation}
where:
$$
s(z)=\frac{(1-\kappa^2)^2}{[1+\kappa^2+2(-1)^{p}\kappa\cos(p\pi z)]^2	},
$$
$$
\kappa=\frac{\sinh(p\tau)}{\sqrt{1+\sinh^{2}(p\tau)}},
$$
$$
\tau=\frac{1}{2L_0}\epsilon\pi t,
$$
being $t=N/p$ and $N$ a non-negative integer. The energy density ${\cal T}^{(a)}$
in Eq. (\ref{T-a}) also can be written in terms of the vacuum and non-vacuum contributions
as ${\cal T}^{(a)}={\cal T}^{(a)}_{\mbox{\footnotesize vac}}+{\cal T}^{(a)}_{\mbox{\footnotesize non-vac}}$, where:
\begin{equation}
{\cal T}^{(a)}_{\mbox{\footnotesize vac}}=(p^2-1)f^{(s)}[s(u)+s(v)]- 2 p^2 f^{(s)},
\label{T-a-vac}
\end{equation}
\begin{equation}
{\cal T}^{(a)}_{\mbox{\footnotesize non-vac}}=-g^{(s)}[s(u)+s(v)].
\label{T-a-non-vac}
\end{equation}
Let us focus initially on the case $p>1$. Since $(p^2-1)f^{(s)}>0$  and  $-g^{(s)}>0$, Eqs. (\ref{T-a-vac}) and (\ref{T-a-non-vac})
give that the functions ${\cal T}^{(a)}_{\mbox{\footnotesize vac}}$ and ${\cal T}^{(a)}_{\mbox{\footnotesize non-vac}}$ have
the same structure \cite{Dodonov-Andreata-JPA-2000}.
To conciliate this conclusion with our first conclusion mentioned above, 
we will show next, starting from our exact approach, that we can find a class of motions for which 
the energy density (\ref{T-total}) is ${\cal T}\approx {\cal T}^{(a)}$,
so that ${\cal T}_{\mbox{\footnotesize vac}}$ and ${\cal T}_{\mbox{\footnotesize non-vac}}$ exhibit approximately the same structure, and that the particular laws of motion 
(\ref{law-of-motion}) - investigated in Ref. \cite{Dodonov-Andreata-JPA-2000} - belong to this class.  

Looking for conditions under which
${\cal T}_{\mbox{\footnotesize vac}}$ and ${\cal T}_{\mbox{\footnotesize non-vac}}$ have the same structure we find that one condition is provided by the laws of motion for which ${\cal{\widetilde{A}}}(z)$ and ${\cal{\widetilde{B}}}(z)$ have a linear relation of the form:
\begin{equation}
{\cal{\widetilde{B}}}(z)=k_{1}{\cal{\widetilde{A}}}(z)+k_{2},
\label{linear-relation}
\end{equation}
where $k_1$ and $k_2$ are constants. From the properties given in Eq. (\ref{A-b-cond}), we get $k_{1}=-k_{2}$, resulting in:
\begin{equation}
{\cal{\widetilde{B}}}(z)=k_{1}\left[{\cal{\widetilde{A}}}(z)-1\right],
\label{linear-relation-new}
\end{equation}
and from Eqs. (\ref{T-vac-A-B}), (\ref{T-non-vac-A}) and (\ref{linear-relation-new}), we have:
\begin{equation}
{\cal T}_{\mbox{\footnotesize vac}}=-\left( f^{(s)}+k_{1} \right)\left[ {\cal{\widetilde{A}}}(u) + {\cal{\widetilde{A}}}(v)\right] + 2k_{1},
\label{01}
\end{equation}
\begin{equation}
{\cal T}_{\mbox{\footnotesize non-vac}}=-g^{(s)}\left[ {\cal{\widetilde{A}}}(u) + {\cal{\widetilde{A}}}(v) \right],
\label{02}
\end{equation}
\begin{equation}
{\cal T}=-\left( h^{(s)}+k_{1} \right)\left[ {\cal{\widetilde{A}}}(u) + {\cal{\widetilde{A}}}(v) \right]+2k_{1}.
\label{03}
\end{equation}
Then, our second conclusion is that 
if the constant factors multiplying ${\cal{\widetilde{A}}}(u) + {\cal{\widetilde{A}}}(v)$ are different from zero and have the same sign, as the ratio
\begin{equation}
\sigma(z)=\frac{{\cal{\widetilde{B}}}(z)}{{\cal{\widetilde{A}}}(z)-1}
\label{sigma}
\end{equation}
becomes more close to a constant value $k_{1}$, that means
\begin{equation}
\sigma(z)\approx k_1,
\label{sigma-k1}
\end{equation}
more the structures of the functions ${\cal T}_{\mbox{\footnotesize vac}}$ and ${\cal T}_{\mbox{\footnotesize non-vac}}$ become similar to one another, whereas if they have different sign then
where we find valleys and peaks in a graph, we can have respectively peaks and valleys in the other.
As a direct consequence, for the class of motions obeying the condition (\ref{sigma-k1}) the quantum radiation forces 
${\cal F}_{\mbox{\footnotesize vac}}$ and ${\cal F}_{\mbox{\footnotesize non-vac}}$
also have similar structures.

Comparing our Eq. (\ref{03}) with (\ref{T-a}), we see that both have the same structure. Our third conclusion is that the formula (\ref{T-a}) found by Andreata and Dodonov belongs to the particular class of formulas (given by Eq. (\ref{03})) for the energy density. We will show that this occurs because for the laws of motion (\ref{law-of-motion}), as the condition $\left\vert \epsilon \right\vert \ll 1$ is better satisfied, better is satisfied the condition (\ref{sigma-k1}). 

In Fig.  1, using (\ref{a-tilde}) and (\ref{b-tilde}), we plot the ratio $\sigma$ (Eq. (\ref{sigma})) for $p=2$, taking into account three values of amplitudes of oscillation: $\epsilon=10^{-3}$, $\epsilon=10^{-2}$ and $\epsilon=10^{-1}$. We observe that $\sigma$ is more approximately the constant value 
$-4 f^{(s)}$ for $\epsilon=10^{-3}$ (dash-dotted line) than for $\epsilon=10^{-1}$ (solid line). 

\begin{figure}[h!]
\begin{center}
\includegraphics[scale=0.7,angle=00]{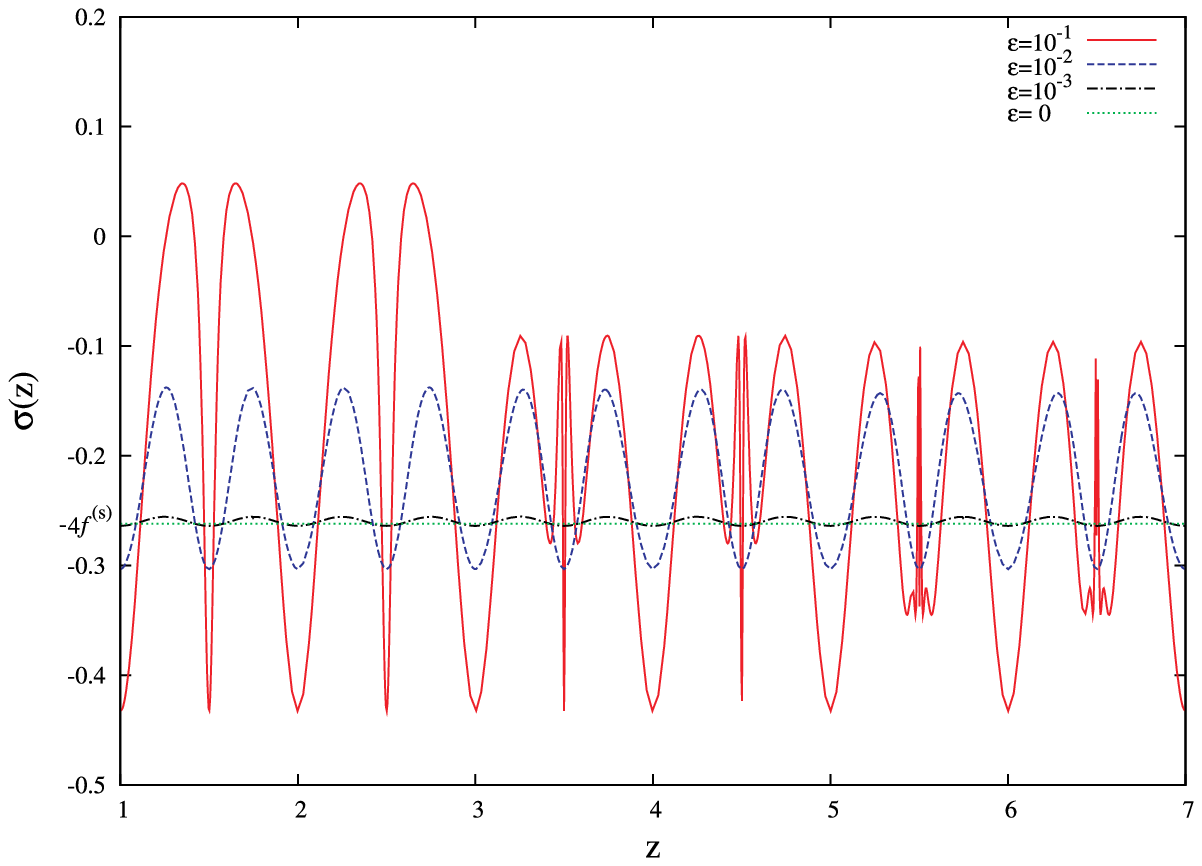}
\caption{The ratio $\sigma(z)={\cal{\widetilde{B}}}(z)/\left[ {\cal{\widetilde{A}}}(z) - 1 \right]$ for the law of motion given by Eq. (\ref{law-of-motion}) with $p=2$. We use different scales for $\sigma(z)$ in each case.
The solid line corresponds to the case $\epsilon=10^{-1}$.
The dashed line corresponds to the case $\epsilon=10^{-2}$, 
exhibiting $40\times[\sigma(z)+4 f^{(s)}]-4f^{(s)}$. The dash-dotted line corresponds to the case $\epsilon=10^{-3}$,
showing $200\times[\sigma(z)+4 f^{(s)}]-4f^{(s)}$. 
The dotted line corresponds to the case $\epsilon=0$.
As $\epsilon\rightarrow 0$ we have $\sigma(z)\rightarrow -4f^{(s)}$.}
\label{sigma-1}
\end{center}
\end{figure}
\begin{figure}[h!]
\begin{center}
\includegraphics[scale=0.7,angle=00]{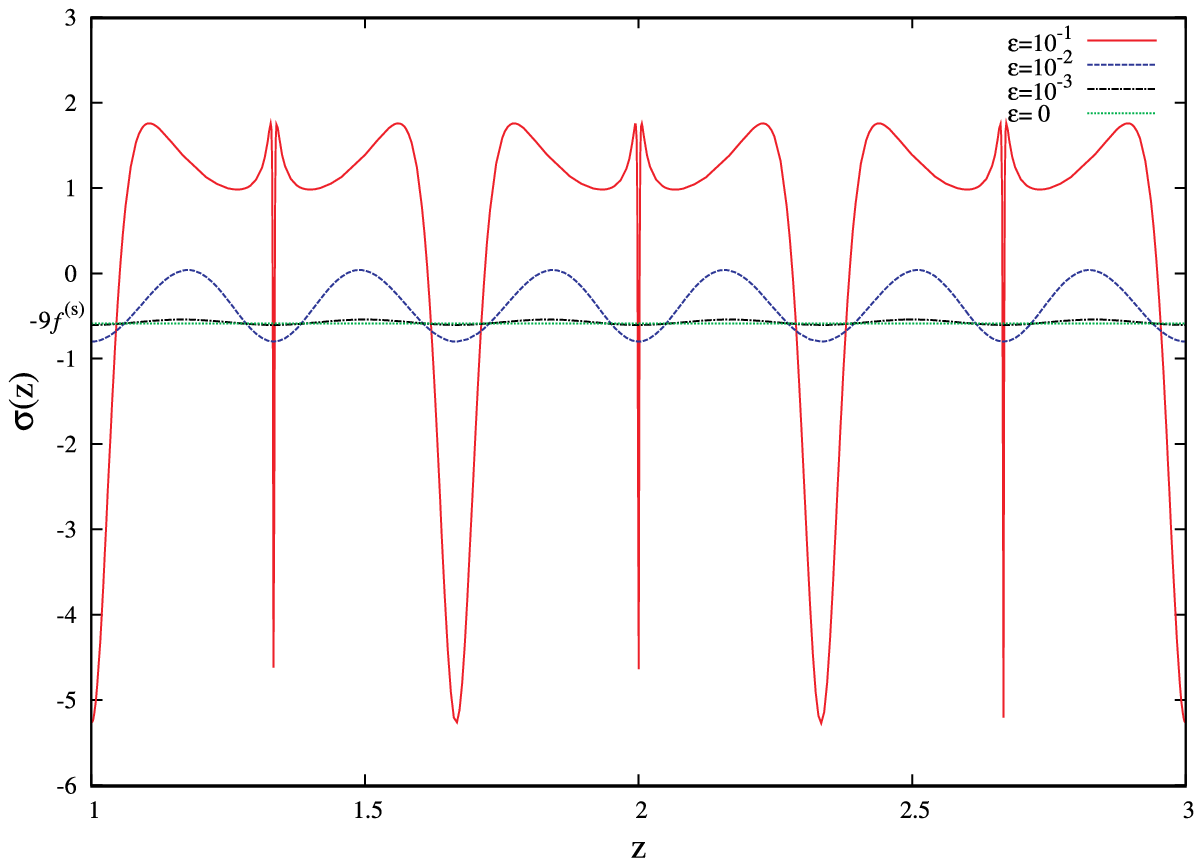}
\caption{The ratio $\sigma(z)={\cal{\widetilde{B}}}(z)/\left[ {\cal{\widetilde{A}}}(z) - 1 \right]$ for the law of motion given by Eq. (\ref{law-of-motion}) with $p=3$. We use different scales for $\sigma(z)$ in each case. 
The solid line corresponds to the case $\epsilon=10^{-1}$.
The dashed line corresponds to the case $\epsilon=10^{-2}$, 
exhibiting $40\times[\sigma(z)+9 f^{(s)}]-9f^{(s)}$. 
The dash-dotted line corresponds to the case $\epsilon=10^{-3}$,
showing $300\times[\sigma(z)+9 f^{(s)}]-9f^{(s)}$. 
The dotted line corresponds to the case $\epsilon=0$.
As $\epsilon\rightarrow 0$ we have $\sigma(z)\rightarrow -9f^{(s)}$.}
\label{sigma-2}
\end{center}
\end{figure}
\begin{figure}[ht!]
\begin{center}
\includegraphics[scale=0.7,angle=00]{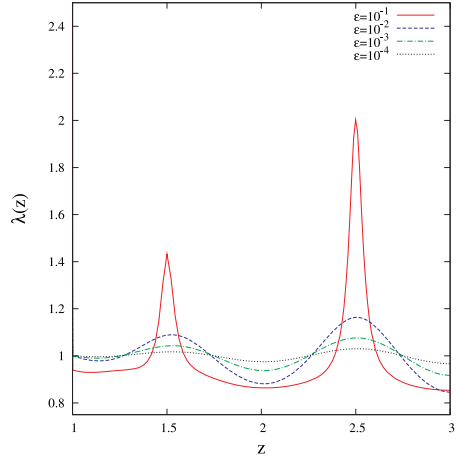}
\caption{The ratio $\lambda(z)={\cal{\widetilde{A}}}(z)/s(z)$ for the law of motion given by Eq. (\ref{law-of-motion}) with $p=2$.
We use different scales for $\lambda(z)$ in each case. The solid line describes the case $\epsilon=10^{-1}$, exhibited as $1/6\times [\lambda(z)-1]+1$.
The dashed line describes the case $\epsilon=10^{-2}$. The dash-dotted line corresponds to the case $\epsilon=10^{-3}$, exhibiting $1/5\times\lambda(z)$, whereas the dotted line corresponds to the case 
$\epsilon=10^{-4}$, showing $20\times\lambda(z)$.
As $\epsilon\rightarrow 0$ we have $\lambda(z)\rightarrow 1$.}
\label{approx}
\end{center}
\end{figure}

In Fig. 2 we plot $\sigma$ for the case $p=3$,
observing that as $\epsilon$ becomes smaller $\sigma$ tends to the value $-9 f^{(s)}$.
Then we see that, in the limit $\epsilon\rightarrow 0$ we get:
\begin{eqnarray}
\sigma\rightarrow k_1 = -p^2 f^{(s)}.
\label{map-1}
\end{eqnarray}

In Fig.  3, we plot the ratio $\lambda(z)={\cal{\widetilde{A}}}(z)/s(z)$ 
for $p=2$, taking into account the values $\epsilon=10^{-4}$, $\epsilon=10^{-3}$, $\epsilon=10^{-2}$ and $\epsilon=10^{-1}$. We observe that 
as $\epsilon\rightarrow 0$ we have $\lambda(z)\rightarrow 1$, what means:

\begin{eqnarray}
{\cal{\widetilde{A}}}(z)\rightarrow s(z).
\label{map-2}
\end{eqnarray}
Eqs. (\ref{map-1}) and (\ref{map-2}) complete the mapping between Eqs. (\ref{03}) and (\ref{T-a}),                                                                 demonstrating a perfect agreement between two completely different approaches to the problem 

Now, let us investigate the following point:
since Eqs. (\ref{T-a-vac}) and (\ref{T-a-non-vac}) are approximations,
we should find differences between the structures of the vacuum
and non-vacuum parts when we consider the case of the law of motion (\ref{law-of-motion}) with our exact approach. To investigate this issue,
we study the law of motion (\ref{law-of-motion}) for $\epsilon=10^{-2}$ and $\epsilon=10^{-1}$. 
Our aim now is to verify, using the exact approach, the similarities and differences between the structures of ${\cal T}_{\mbox{\footnotesize vac}}$ and  ${\cal T}_{\mbox{\footnotesize non-vac}}$ 
for both values of $\epsilon$. 

Although the formulas (\ref{a-tilde}), (\ref{b-tilde}), (\ref{T-vac-A-B}) and (\ref{T-non-vac-A}) are formally exact,
to extract numerical values for ${\cal T}_{\mbox{\footnotesize vac}}$ and  
${\cal T}_{\mbox{\footnotesize non-vac}}$ we need to calculate 
the functions $t_i(z)$, which are given by Eqs. (\ref{z1}) and (\ref{z}).
For the law of motion (\ref{law-of-motion}) (and in general) we can get only numerical solutions of 
(\ref{z1}) and (\ref{z}). For a given value of $z$, the Eq. (\ref{z1}) can be solved numerically 
and the result for $t_1$ naturally has a certain limited accuracy. When $t_1$ is used in Eq. (\ref{z})
to calculate $t_2$, the solution of $t_{2}+L(t_{2})=t_1-L(t_1)$ can give a result less accurate
than the result previously obtained for $t_1$, and successive calculations of the remaining values of $t_i$ via equations (\ref{z}) (for 
$i=2,3,4...$) could generate a final result for $t_n$
with a poor accuracy, if compared to the accuracy of the initial value for $t_1$. 
Moreover, when we insert the numerical values 
for $t_i$ in Eqs. (\ref{a-tilde}), (\ref{b-tilde}),
(\ref{T-vac-A-B}) and (\ref{T-non-vac-A}),
the final values for ${\cal T}_{\mbox{\footnotesize vac}}$ and  
${\cal T}_{\mbox{\footnotesize non-vac}}$
could have their accuracy diminished even more.
To deal with this question and control the final accuracy of our results,
we perform the numerical calculations in Maple computer algebra system \cite{maple},
which enables us to control the number of digits used when calculating with floating-point numbers. 
To obtain a final numerical value for ${\cal T}_{\mbox{\footnotesize vac}}$ or ${\cal T}_{\mbox{\footnotesize non-vac}}$,
we carry out several independent calculations using our routines developed in Maple \cite{alves-granhen-CPC-2010}. 
In each calculation we take all numerical solutions
with a certain number of digits. Considering 
from 3 to 100 digits, we observe the convergence of the results as the number of digits 
(related to the initial accuracy considered for the solution of (\ref{z1})) is enhanced. 
This enable us to point which are the exact digits (the accuracy) in our results. For instance,
the values of ${\cal T}_{\mbox{\footnotesize vac}}(10,0.5)$ ($L_0=1,p=2$) performed with 3, 4, 5, 6, 10 and 20 digits are given, respectively by:
0.856, 0.8638, 0.86364, 0.863704, 0.8637005768 and 0.86370057587773139184. Analyzing also the convergence of the 
results up to 100 digits, we can
obtain a final accuracy of 10 or more digits, but we 
just display the result as ${\cal T}_{\mbox{\footnotesize vac}}(10,0.5)\approx 0.864$, where the first two digits can be considered as exact digits.
Hereafter, the exhibited results have accuracy at least up to the penultimate digit shown.

\begin{figure}[h!]
\begin{center}
\includegraphics[scale=0.7, angle=00]{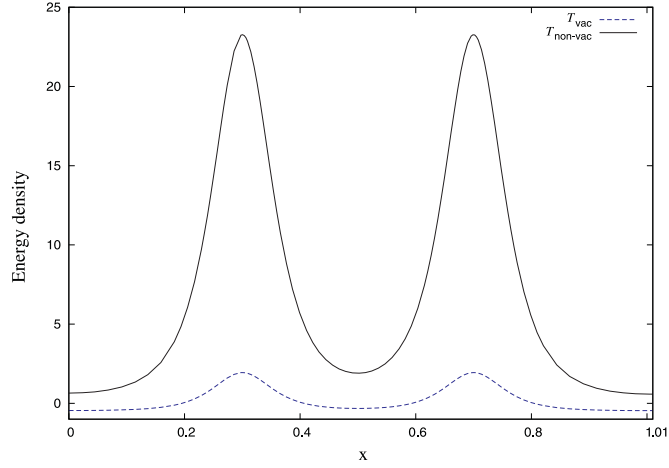}
\caption{The energy densities ${\cal T}_{\mbox{\footnotesize vac}}$ (dashed line) and ${\cal T}_{\mbox{\footnotesize non-vac}}$ (solid line) at the instant $t=20.2$ with $T=1$, plotted via the exact formulas (\ref{T-vac-A-B}) and  (\ref{T-non-vac-A}). We consider the law of motion (\ref{law-of-motion}) with $p=2$ and $\epsilon=10^{-2}$.}
\label{T-vac-non-vac-0-01}
\end{center}
\end{figure}
In Fig. 4, using the formulas (\ref{T-vac-A-B}) and (\ref{T-non-vac-A}) we plot the energy densities 
${\cal T}_{\mbox{\footnotesize vac}}$ and ${\cal T}_{\mbox{\footnotesize non-vac}}$ for the case $p=2$, $T=1$, $\epsilon=10^{-2}$ at $t=20.2$.
We see that both energy densities have two peaks 
(in this case, located at $x=0.30$ and $x=0.70$, with values 
${\cal T}_{\mbox{\footnotesize vac}}(20.2,0.30)={\cal T}_{\mbox{\footnotesize vac}}(20.2,0.70)\approx 1.94$
and ${\cal T}_{\mbox{\footnotesize non-vac}}(20.2,0.30)={\cal T}_{\mbox{\footnotesize non-vac}}(20.2,0.70)$$\approx 23.3$),
both have three minimum points (located at $x=0$,  $x=0.50$, $x=1.01$, with values 
${\cal T}_{\mbox{\footnotesize vac}}(20.2,0)$$\approx-0.453, {\cal T}_{\mbox{\footnotesize vac}}(20.2,0.5	)\approx -0.322$,
${\cal T}_{\mbox{\footnotesize vac}}(20.2,1.01)$$\approx -0.461$,
${\cal T}_{\mbox{\footnotesize non-vac}}(20.2,0)$$\approx 0.647, {\cal T}_{\mbox{\footnotesize non-vac}}(20.2,0.5)$
$\approx 1.9$,
${\cal T}_{\mbox{\footnotesize non-vac}}(20.2,1.01) \approx 0.579$),
so that ${\cal T}_{\mbox{\footnotesize vac}}$ and ${\cal T}_{\mbox{\footnotesize non-vac}}$
exhibit the same structure, as predicted by the approximate analytical formulas (\ref{T-a-vac}) and (\ref{T-a-non-vac}) \cite{Dodonov-Andreata-JPA-2000}, 
which are based on the assumption $\left\vert \epsilon \right\vert \ll 1$. We remark that $x\approx1.01$ corresponds to the position
of the right mirror when $t=20.2$ ($L(20.2)\approx 1.01$).

For the case $\epsilon=10^{-1}$, we get for ${\cal T}_{\mbox{\footnotesize vac}}(20.2,x)$ and ${\cal T}_{\mbox{\footnotesize non-vac}}(20.2,x)$
that both present again two narrow peaks, located at $x=0.30$ and $x=0.70$ (similarly to the case showed in Fig. 4, but with values 
${\cal T}_{\mbox{\footnotesize vac}}(20.2,0.30)={\cal T}_{\mbox{\footnotesize vac}}(20.2,0.70)\approx 0.249\times 10^{13}$,
${\cal T}_{\mbox{\footnotesize non-vac}}(20.2,0.30)={\cal T}_{\mbox{\footnotesize non-vac}}(20.2,0.70)\approx 0.126\times 10^{14}$).
Moreover, in the case $\epsilon=10^{-1}$,
${\cal T}_{\mbox{\footnotesize vac}}(20.2,x)$
exhibits several other maximum and minimum points that are not visualized in the graph of ${\cal T}_{\mbox{\footnotesize non-vac}}(20.2,x)$. 

\begin{figure}[ht!]
\begin{center}
\includegraphics[scale=0.7, angle=00]{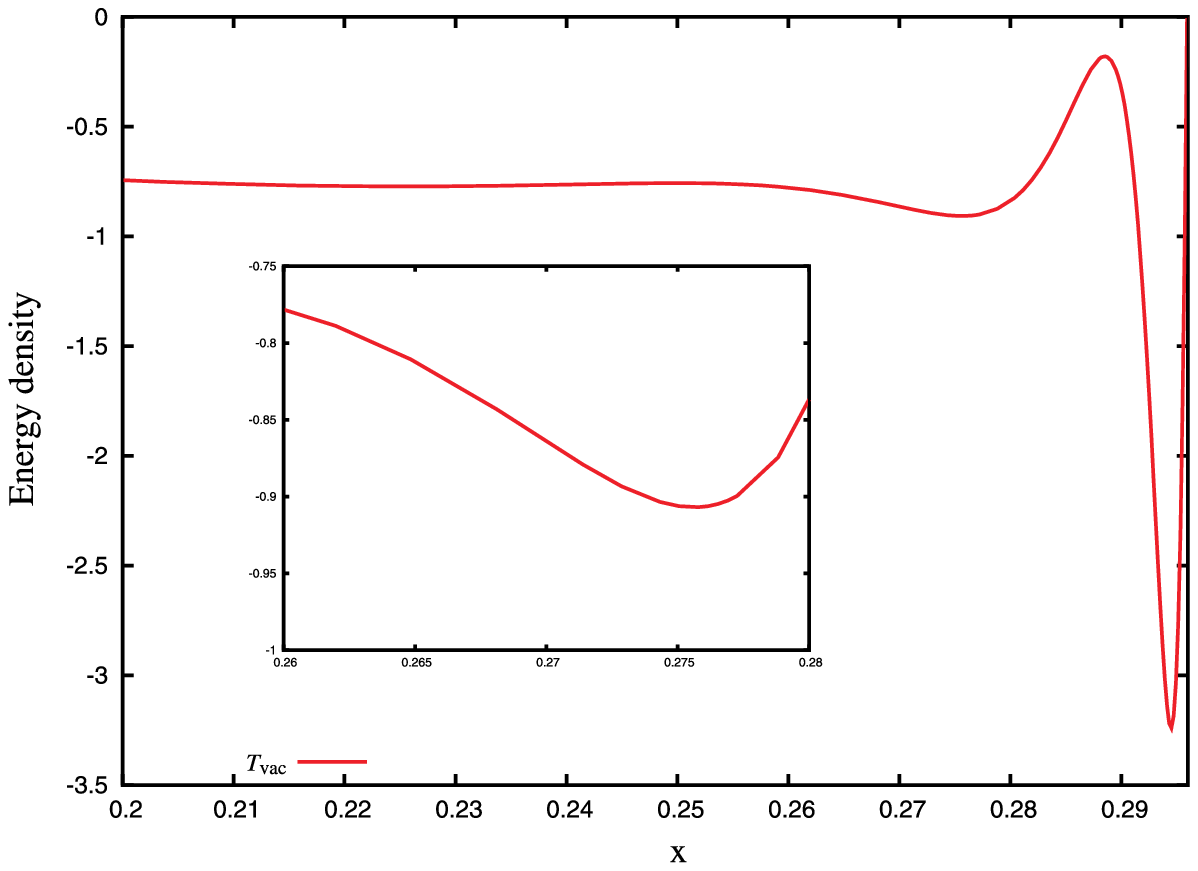}
\caption{Detail of the energy density ${\cal T}_{\mbox{\footnotesize vac}}$ 
at the instant $t=20.2$, for the law of motion (\ref{law-of-motion}) with $p=2$, $T=1$ and $\epsilon=10^{-1}$, plotted 
via the exact formula (\ref{T-vac-A-B}) in the region $0.2<x<0.3$. In detail we show ${\cal T}_{\mbox{\footnotesize vac}}$
in the sub-region $0.26<x<0.28$.  The spacing used between the calculated points in the graph is $10^{-4}$.
}
\label{T-vac-non-vac-0-1-parte-1}
\end{center}
\end{figure}
\begin{figure}[ht!]
\begin{center}
\includegraphics[scale=0.7, angle=00]{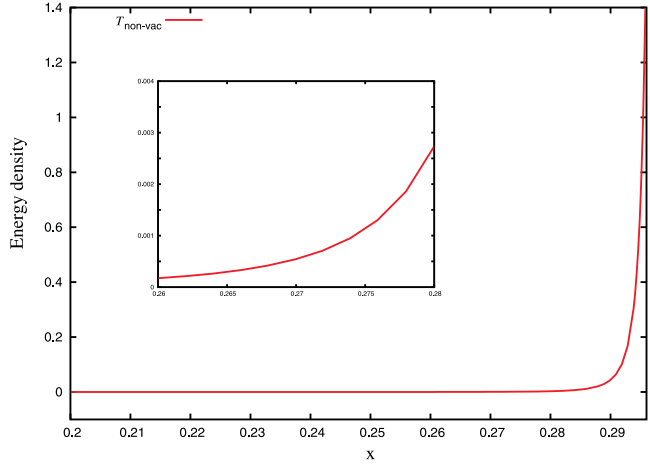}
\caption{The energy density ${\cal T}_{\mbox{\footnotesize non-vac}}$ 
at the instant $t=20.2$, for the law of motion (\ref{law-of-motion}) with $p=2$ and $\epsilon=10^{-1}$, plotted 
via the exact formula (\ref{T-non-vac-A}) in the region $0.2<x<0.3$. In detail we show ${\cal T}_{\mbox{\footnotesize non-vac}}$
in the sub-region $0.26<x<0.28$.  The spacing used between the calculated points in the graph is $10^{-4}$.}
\label{T-vac-non-vac-0-1-parte-2}
\end{center}
\end{figure}

In Fig. 5 we investigate details of the behavior of ${\cal T}_{\mbox{\footnotesize vac}}$ for the case $\epsilon=10^{-1}$ in the region $0.2<x<0.3$. We can see peaks for the following values: $x\approx 0.250$ and $x\approx 0.289$; we also see minimum values at the points $x\approx 0.230$, $x\approx 0.28$ and $x\approx 0.295$. 

In Fig. 6 we see the behavior of ${\cal T}_{\mbox{\footnotesize non-vac}}$ (for $\epsilon=10^{-1}$ and $T=1$), in the same region as Fig. 5, but
there is no peak or valley.
Then we verify that when we consider $\epsilon=10^{-2}$ and $\epsilon=10^{-1}$, since the
the former value is in better agreement with the conditions $\left\vert \epsilon \right\vert \ll 1$ and
$\sigma\approx-p^2f^{(s)}= -4\pi$, no difference
between the structures of ${\cal T}_{\mbox{\footnotesize vac}}$ and ${\cal T}_{\mbox{\footnotesize non-vac}}$
is perceived, but for the latter value of $\epsilon$
differences between the structures of ${\cal T}_{\mbox{\footnotesize vac}}$ (Fig. 6) and ${\cal T}_{\mbox{\footnotesize non-vac}}$ (Fig. 5) become 
evident, as predicted via the exact formulas (\ref{T-vac-A-B}) and (\ref{T-non-vac-A}).

\begin{figure}[ht!]
\begin{center}
\includegraphics[scale=0.7, angle=00]{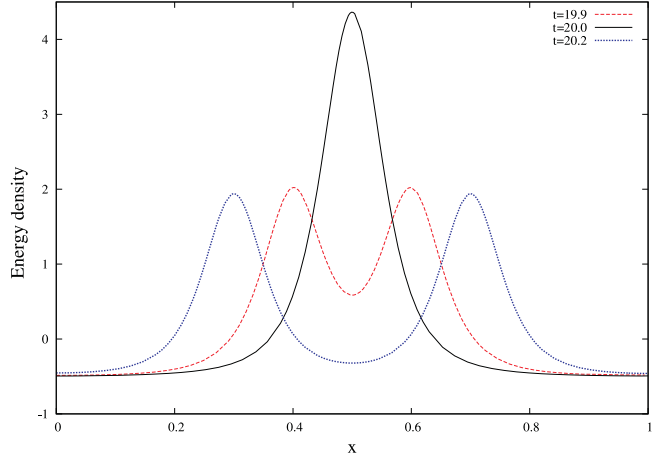}
\caption{The time evolution of the energy density ${\cal T}_{\mbox{\footnotesize vac}}$ for 
the law of motion (\ref{law-of-motion}), with $p=2$ and $\epsilon=10^{-2}$, plotted 
via the exact formula (\ref{T-vac-A-B}). 
The dashed line shows two peaks at $t=19.9$, initializing the merging process. The solid line shows the maximum peak formed at 
$x=0.5$ and $t=20$. The dotted line shows the energy density at $t=20.2$, after the merging.}
\label{eps-001}
\end{center}
\end{figure}

For $p=1$, Eq. (\ref{T-a-vac}) gives ${\cal T}_{\mbox{\footnotesize vac}}\approx{\cal T}^{(a)}_{\mbox{\footnotesize vac}}=- 2 f^{(s)}={\cal T}_{\mbox{\footnotesize cas}}$,
so that the energy density would conserve its vacuum (Casimir) value, whereas Eq. (\ref{T-a-non-vac}) gives 
for ${\cal T}^{(a)}_{\mbox{\footnotesize non-vac}}$ a spacetime dependence. 
However, the exact behavior is given by Eqs. (\ref{T-vac-A-B}) and (\ref{T-non-vac-A}), which show
that the values of both ${\cal T}_{\mbox{\footnotesize vac}}$ and ${\cal T}_{\mbox{\footnotesize non-vac}}$  change in time and space.
For instance, for the law of motion (\ref{law-of-motion}) with $p=1$ and $\epsilon=10^{-2}$,
the exact behavior of ${\cal T}^{(a)}_{\mbox{\footnotesize vac}}(100.5,x)$ exhibits two minimum points surrounding a peak
located at $x=0.5$ (see also Ref. \cite{Cole-Schieve-1995}), whereas
${\cal T}^{(a)}_{\mbox{\footnotesize non-vac}}(100.5,x)$ displays a peak at $x=0.5$. For this case $p=1$, Eq. (\ref{map-1}) remains valid, so that
as $\epsilon\rightarrow 0$, ${\cal T}^{(a)}_{\mbox{\footnotesize vac}}$ and ${\cal T}^{(a)}_{\mbox{\footnotesize non-vac}}$
display more similar structures.

Now, we will compare exact and approximate results in the prediction of the maximum value of the peaks in the energy density moving in an
oscillating cavity.
In this context, let us analyze again the behavior of ${\cal T}_{\mbox{\footnotesize vac}}(t,x)$ 
and ${\cal T}_{\mbox{\footnotesize non-vac}}(t,x)$ for the case $p=2$.
The two peaks showed in Fig. 4 (for ${\cal T}_{\mbox{\footnotesize vac}}(t,x)$ 
or ${\cal T}_{\mbox{\footnotesize non-vac}}(t,x)$) have
the same value (see values presented above), move in opposite direction and at $(t,x)=(N,0.5)$ ($N$ is a non-negative integer) they merge forming a single maximum peak (see Fig. 7).
The maximum value of the energy density occurs when the two peaks merge. This value is represented 
for vacuum and non-vacuum parts, respectively, by
${\cal T}_{\mbox{\footnotesize vac}}^{{\mbox{\footnotesize max}}}$
and 
${\cal T}_{\mbox{\footnotesize non-vac}}^{{\mbox{\footnotesize max}}}$.
Exactly, we have 
\begin{equation}
{\cal T}_{\mbox{\footnotesize vac}}^{{\mbox{\footnotesize max}}}={\cal T}_{\mbox{\footnotesize vac}}(N,0.5),
\label{T-max-vac}
\end{equation}
\begin{equation}
{\cal T}_{\mbox{\footnotesize non-vac}}^{{\mbox{\footnotesize max}}}={\cal T}_{\mbox{\footnotesize non-vac}}(N,0.5).
\label{T-max-non-vac}
\end{equation}
From an approximate analysis, taking as basis the results found in Ref. \cite{Dodonov-Andreata-JPA-2000},
${\cal T}_{\mbox{\footnotesize vac}}^{{\mbox{\footnotesize max}}}$ and ${\cal T}_{\mbox{\footnotesize vac}}^{{\mbox{\footnotesize max}}}$
grow in time according
to ${\cal T}_{\mbox{\footnotesize vac}}^{{\mbox{\footnotesize max}}}\approx 
{\cal T}_{\mbox{\footnotesize vac}}^{(a){\mbox{\footnotesize max}}}$ and
${\cal T}_{\mbox{\footnotesize non-vac}}^{{\mbox{\footnotesize max}}}\approx 
{\cal T}_{\mbox{\footnotesize non-vac}}^{(a){\mbox{\footnotesize max}}}$, where 
\begin{equation}
{\cal T}_{\mbox{\footnotesize vac}}^{(a){\mbox{\footnotesize max}}}=6f^{(s)}(e^{8\tau}-1)-2f^{(s)},
\label{T-max-a}
\end{equation}
\begin{equation}
{\cal T}_{\mbox{\footnotesize non-vac}}^{(a){\mbox{\footnotesize max}}}=-2 g^{(s)}\frac{(1+\kappa)^2}{(1-\kappa)^2}.
\label{T-max-a-non-vac}
\end{equation}

In Fig. 8 we examine the case given by the law of motion (\ref{law-of-motion}), $p=2$ and $\epsilon=10^{-2}$,
and visualize agreement between the growing in time of the peaks predicted by the approximate formula (\ref{T-max-a})
and the exact values obtained via (\ref{T-max-vac}). In Table 1, again considering $\epsilon=10^{-2}$, 
we compare ${\cal T}_{\mbox{\footnotesize vac}}^{{\mbox{\footnotesize max}}}$ with ${\cal T}_{\mbox{\footnotesize vac}}^{(a){\mbox{\footnotesize max}}}$ for larger times, and also compare ${\cal T}_{\mbox{\footnotesize non-vac}}^{{\mbox{\footnotesize max}}}$ with ${\cal T}_{\mbox{\footnotesize non-vac}}^{(a){\mbox{\footnotesize max}}}$. We remark the increasing error of the approximate formulas in comparison with the exact formulas,
observing that the error of ${\cal T}_{\mbox{\footnotesize non-vac}}^{(a){\mbox{\footnotesize max}}}$ grows more rapidly.

\begin{figure}[ht!]
\begin{center}
\includegraphics[scale=0.7, angle=00]{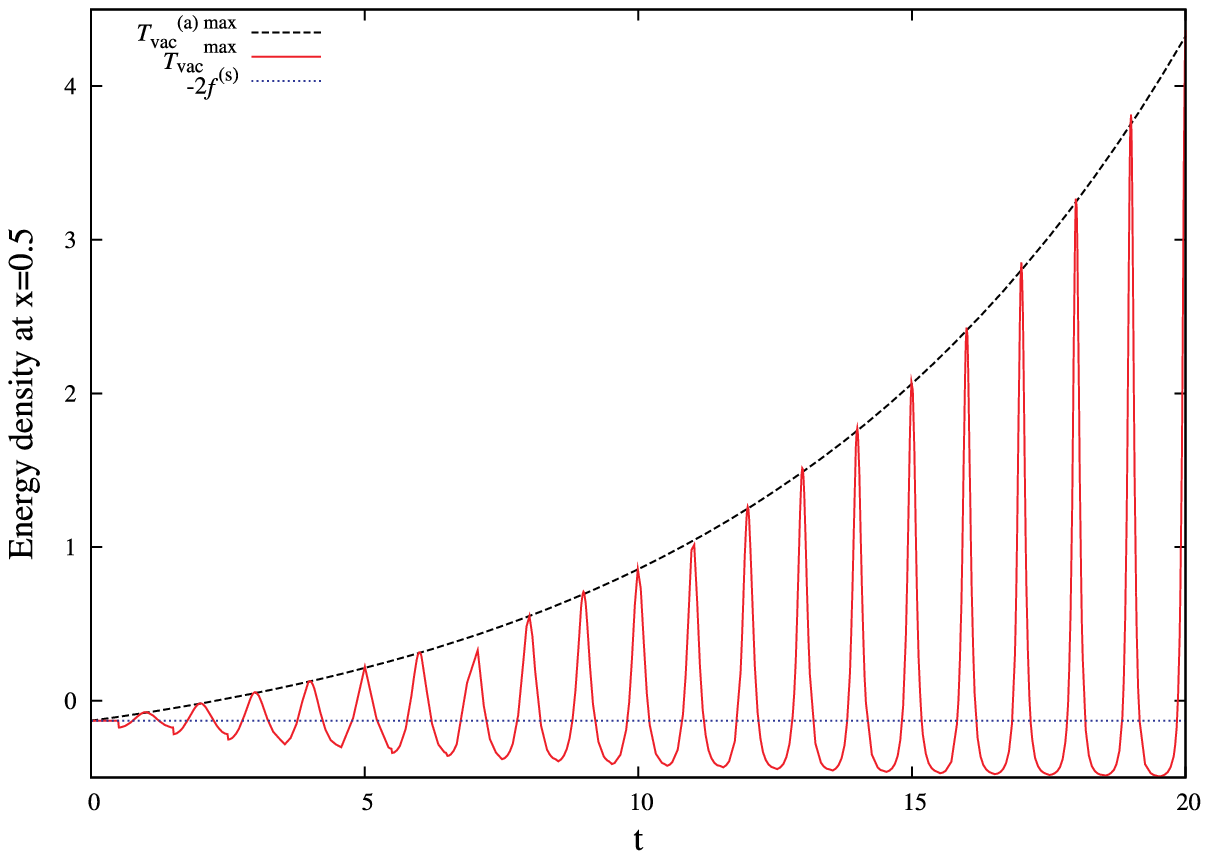}
\caption{The energy density ${\cal T}_{\mbox{\footnotesize vac}}^{{\mbox{\footnotesize max}}}$ (solid line),
given by Eq. (\ref{T-max-vac}), and its growing in
time predicted by (\ref{T-max-a}), for the case of $\epsilon=10^{-2}$.}
\label{picos-1}
\end{center}
\end{figure}

\begin{table}
\caption{${\cal T}_{\mbox{\footnotesize vac}}^{{\mbox{\footnotesize max}}}$ 
and ${\cal T}_{\mbox{\footnotesize non-vac}}^{{\mbox{\footnotesize max}}}$ (for $T=1$)  
computed via numerical exact method (column 2) and via approximate analytical formula (column 3), 
with $L_0=1$, $\epsilon=10^{-2}$ and $p=2$. The percent error 
$({\cal T}_{\mbox{\footnotesize [vac, non-vac]}}^{{\mbox{\footnotesize max}}}-{\cal T}_{\mbox{\footnotesize [vac, non-vac]}}^{(a){\mbox{\footnotesize max}}})/
{\cal T}_{\mbox{\footnotesize [vac, non-vac]}}^{{\mbox{\footnotesize max}}}\times 100$
is showed in column 4.}
\begin{center}
\begin{tabular}{|c|c|c|c|}
  \hline	
  {\footnotesize \mbox{${t}$}} & {\footnotesize ${\cal T}_{\mbox{\footnotesize vac}}^{{\mbox{\footnotesize max}}}$} & {\footnotesize ${\cal T}_{\mbox{\footnotesize vac}}^{(a){\mbox{\footnotesize max}}}$} &{\footnotesize Percent Error}
   \\
  \hline
 {\footnotesize \mbox{$10$}} & {\footnotesize $0.864$ } & {\footnotesize $0.856$ } 
  &{\footnotesize $0.870$}
   \\
  \hline
 {\footnotesize \mbox{$10^2$}} & {\footnotesize $115$ } & {\footnotesize $113$ } 
  &{\footnotesize $2.16$} 
   \\
  \hline
 {\footnotesize \mbox{$5\times 10^2$}} & {\footnotesize $0.832\times 10^{27}$ } & {\footnotesize $0.761\times 10^{27}$ } 
  &{\footnotesize $8.44$} 
  \\
  \hline
 {\footnotesize \mbox{$10^3$}} & {\footnotesize $0.175\times 10^{55}$ } & {\footnotesize $0.148\times 10^{55}$ } 
  &{\footnotesize $15.7$} 
  \\
  \hline
    \hline	
  {\footnotesize \mbox{${}$}} & {\footnotesize ${\cal T}_{\mbox{\footnotesize non-vac}}^{{\mbox{\footnotesize max}}}$} & {\footnotesize ${\cal T}_{\mbox{\footnotesize non-vac}}^{(a){\mbox{\footnotesize max}}}$} &{\footnotesize }
   \\
  \hline
 {\footnotesize \mbox{$10$}} & {\footnotesize $13.12$ } & {\footnotesize $13.1$ } 
  &{\footnotesize $0.17$}
   \\
  \hline
 {\footnotesize \mbox{$10^2$}} & {\footnotesize $109\times10^4$ } & {\footnotesize $107\times10^4$ } 
  &{\footnotesize $1.64$} 
   \\
  \hline
 {\footnotesize \mbox{$5\times 10^2$}} & {\footnotesize $0.785\times 10^{28}$ } & {\footnotesize $0.723\times 10^{28}$ } 
  &{\footnotesize $7.95$} 
  \\
  \hline
 {\footnotesize \mbox{$10^3$}} & {\footnotesize $0.165\times 10^{56}$ } & {\footnotesize $0.149\times 10^{42}$ } 
  &{\footnotesize $99.9$} 
  \\
  \hline
\end{tabular}
\end{center}
\label{tabela-vac-0-01}
\end{table}

\vspace{10cm}

In Fig. 9 we examine the case discussed in Fig. 8, but with a larger amplitude: $\epsilon=10^{-1}$. Now,
we see disagreement between the maximum value of the energy density (\ref{T-max-vac}) and its growing in
time predicted by (\ref{T-max-a}). In Table 2, we examine the case discussed in Fig. 9, but for larger instants.
We see large discrepancy between approximate and exact formulas. 

\begin{figure}[t]
\begin{center}
\includegraphics[scale=0.7, angle=00]{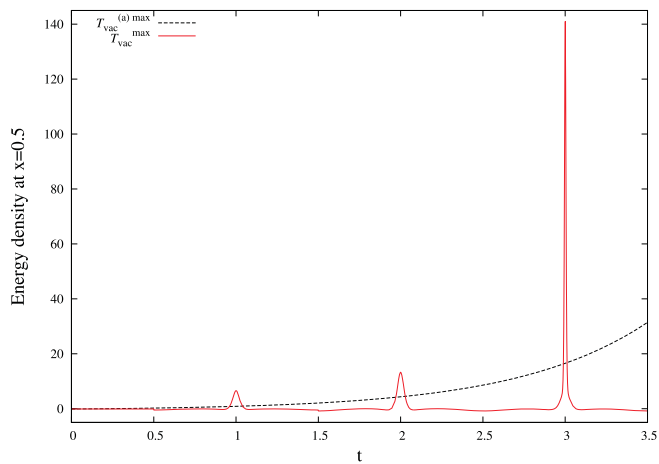}
\caption{The energy density ${\cal T}_{\mbox{\footnotesize vac}}^{{\mbox{\footnotesize max}}}$ (solid line),
given by Eq. (\ref{T-max-vac}), and its growing in
time predicted by (\ref{T-max-a}), for the case of $\epsilon=10^{-1}$.}
\label{picos-2}
\end{center}
\end{figure}

\begin{table}
\caption{${\cal T}_{\mbox{\footnotesize vac}}^{{\mbox{\footnotesize max}}}$ and ${\cal T}_{\mbox{\footnotesize non-vac}}^{{\mbox{\footnotesize max}}}$
(for $T=1$) computed via numerical exact method (column 2) and via approximate analytical formula (column 3), 
with $L_0=1$, $\epsilon=10^{-1}$ and $p=2$. The percent error 
$({\cal T}_{\mbox{\footnotesize [vac, non-vac]}}^{{\mbox{\footnotesize max}}}-{\cal T}_{\mbox{\footnotesize [vac, non-vac]}}^{(a){\mbox{\footnotesize max}}})/
{\cal T}_{\mbox{\footnotesize [vac, non-vac]}}^{{\mbox{\footnotesize max}}}\times 100$
is showed in column 4.}
\begin{center}
\begin{tabular}{|c|c|c|c|}
  \hline	
  {\footnotesize \mbox{${t}$}} & {\footnotesize ${\cal T}_{\mbox{\footnotesize vac}}^{{\mbox{\footnotesize max}}}$} & {\footnotesize ${\cal T}_{\mbox{\footnotesize vac}}^{(a){\mbox{\footnotesize max}}}$} &{\footnotesize Percent Error}
   \\
  \hline
 {\footnotesize \mbox{$10$}} & {\footnotesize $191\times10^4$ } & {\footnotesize $113\times10^3$ } 
  &{\footnotesize $94.1$} 
   \\
  \hline
 {\footnotesize \mbox{$40$}} & {\footnotesize $0.338\times 10^{26}$ } & {\footnotesize $0.266\times 10^{22}$ } 
  &{\footnotesize $99.9$} 
  \\
  \hline
 \hline	
  {\footnotesize \mbox{${}$}} & {\footnotesize ${\cal T}_{\mbox{\footnotesize non-vac}}^{{\mbox{\footnotesize max}}}$} & {\footnotesize ${\cal T}_{\mbox{\footnotesize non-vac}}^{(a){\mbox{\footnotesize max}}}$} &{\footnotesize }
   \\
  \hline
 {\footnotesize \mbox{$10$}} & {\footnotesize $971\times10^4$ } & {\footnotesize $107\times10^4$ } 
  &{\footnotesize $88.9$} 
   \\
  \hline
 {\footnotesize \mbox{$40$}} & {\footnotesize $0.171\times 10^{27}$ } & {\footnotesize $0.252\times 10^{23}$ } 
  &{\footnotesize $99.9$} 
  \\
  \hline
  \end{tabular}
\end{center}
\label{tabela-vac-0-1}
\end{table}

In counterpart, if we diminish the amplitude to $\epsilon=10^{-8}$, we get 
an excellent agreement between approximate and exact formulas for short and long times. For instance:
${\cal T}_{\mbox{\footnotesize vac}}^{{\mbox{\footnotesize max}}}(4\times 10^4)\approx$
${\cal T}_{\mbox{\footnotesize vac}}^{(a){\mbox{\footnotesize max}}}(4\times 10^4)\approx -0.129$;
${\cal T}_{\mbox{\footnotesize non-vac}}^{{\mbox{\footnotesize max}}}(4\times 10^4)\approx$
${\cal T}_{\mbox{\footnotesize non-vac}}^{(a){\mbox{\footnotesize max}}}(4\times 10^4)\approx 3.75$.

\vspace{10cm}

\section{Conclusions}
\label{conclusions}

Considering a thermal bath as the initial field state
in a cavity with a moving mirror in a two-dimensional spacetime,
we found for small amplitude of oscillation good agreement between the exact and approximate results for the maximum
values of the energy densities. This agreement
strengthens the validity of the analytical approximate results obtained in Ref. 
\cite{Dodonov-Andreata-JPA-2000}, and also reinforce the validity of the exact formulas and numerical results 
discussed here.  
However, for larger values of $\epsilon$, as shown in Table \ref{tabela-vac-0-1}, significant 
discrepancies appear.
This is expected, since
the analytical formulas are valid for $\left\vert \epsilon \right\vert \ll 1$.
Then, we see that the exact formulas (\ref{T-vac-A-B}) and (\ref{T-non-vac-A}) can give results for cases of large amplitudes, which are out of reach of the perturbative approaches found in the literature. 

We showed that the energy densities ${\cal T}_{\mbox{\footnotesize vac}}$ and ${\cal T}_{\mbox{\footnotesize non-vac}}$ 
have, in general, different structures. However, we found that these energy densities
can exhibit approximately the same structure for 
a class of laws of motion
for which
the ratio given in Eq. (\ref{sigma}) is approximately a constant value.
We also showed that this condition is just satisfied by the oscillating laws of motion with small amplitude investigated in 
the literature, specifically in Ref. \cite{Dodonov-Andreata-JPA-2000}, where
the same structure for these energy densities was predicted via approximate methods. 
We verified that for this class of laws of motion there is a
direct mapping between the approximate analytical formulas for the energy density
found in the literature and the exact formulas discussed here. 
On the other hand, we found that for larger amplitudes of oscillation 
the ratio in Eq. (\ref{sigma}) becomes far from  a constant value, displaying larger oscillatory behavior. This means that
that ${\cal T}_{\mbox{\footnotesize vac}}$ and ${\cal T}_{\mbox{\footnotesize non-vac}}$ can display
different structures. Moreover, the exact formulas (\ref{T-vac-A-B}) and (\ref{T-non-vac-A}) 
can say precisely how these structures are.
Finally, we remark that, beyond the thermal case, the conclusions found in our letter are directly extensible
to any other initial state whose density matrix is diagonal in the Fock basis.

\section*{Acknowledgements}

We acknowledge the Referees for many constructive criticisms and suggestions to improve the final version of this letter. 
H.O.S. acknowledges the hospitality of Instituto de F\'isica-UFRJ where part of this work was done. This work was supported by CNPq, CAPES and FAPESPA - Brazil.

\section*{References}
\bibliographystyle{elsarticle-num}
%

\end{document}